# The Proposed High Energy Telescope (HET) for *EXIST*

J. Hong[a][1], J. Grindlay[a], B. Allen[a], G. Skinner[b], S. Barthelmy[b], N. Gehrels[b],
A. Garson[c], H. Krawczynski[c], W. Cook[d], F. Harrison[d],
L. Natalucci[e], P. Ubertini[e] and the *EXIST*/HET team

[a] Harvard Smithsonian Center for Astrophysics, Cambridge, MA 02138;
[b] NASA Goddard Space Flight Center, Greenbelt, MD 20771;
[c] Washington University at St. Louis, St. Louis, MO 63130;
[d] California Institute of Technology, Pasadena, CA 91125;
[e] Istituto di Astrofisica Spaziale e Fisica Cosmica, Rome, Italy

**ABSTRACT**

The hard X-ray sky now being studied by *INTEGRAL* and *Swift* and soon by *NuSTAR* is rich with energetic phenomena and highly variable non-thermal phenomena on a broad range of timescales. The High Energy Telescope (HET) on the proposed Energetic X-ray Imaging Survey Telescope (*EXIST*) mission will repeatedly survey the full sky for rare and luminous hard X-ray phenomena at unprecedented sensitivities. It will detect and localize (<20″, at 5σ threshold) X-ray sources quickly for immediate followup identification by two other onboard telescopes - the Soft X-ray imager (SXI) and Optical/Infrared Telescope (IRT). The large array (4.5 m$^2$) of imaging (0.6 mm pixel) CZT detectors in the HET, a coded-aperture telescope, will provide unprecedented high sensitivity (~0.06 mCrab Full Sky in a 2 year continuous scanning survey) in the 5 – 600 keV band. The large field of view (90º × 70º) and zenith scanning with alternating-orbital nodding motion planned for the first 2 years of the mission will enable nearly continuous monitoring of the full sky. A 3y followup pointed mission phase provides deep UV-Optical-IR-Soft X-ray and Hard X-ray imaging and spectroscopy for thousands of sources discovered in the Survey. We review the HET design concept and report the recent progress of the CZT detector development, which is underway through a series of balloon-borne wide-field hard X-ray telescope experiments, *ProtoEXIST*. We carried out a successful flight of the first generation of fine pixel large area CZT detectors (*ProtoEXIST1*) on Oct 9, 2009. We also summarize our future plan (*ProtoEXIST2 & 3*) for the technology development needed for the HET.

## 1. INTRODUCTION

The proposed Energetic X-ray Imaging Survey Telescope (*EXIST*) mission will explore the early Universe through high redshift Gamma-ray Bursts (GRBs) [1]. Consisting of three complementary telescopes - the High Energy Telescope (HET), the Soft X-ray Imager (SXI) and the Optical/Infrared Telescope (IRT), *EXIST* will serve as a next generation multi-wavelength observatory (Fig. 1). The HET will locate GRBs and transients within 20″ in less than 10 sec after trigger, allowing rapid re-pointing of the observatory for immediate followup SXI and IRT observations for source identification and further study. The HET will also survey the hard X-ray sky down to 0.06 mCrab during the first two years of the scanning operation, which will discover black holes on all scales from stellar mass to supermassive center of Active Galactic Nuclei (AGN). Continuous scan of the entire sky every two orbits is ideal for monitoring variable X-ray sources and for capturing rare events such as soft Gamma-ray repeaters and tidal disruption events of stars by dormant

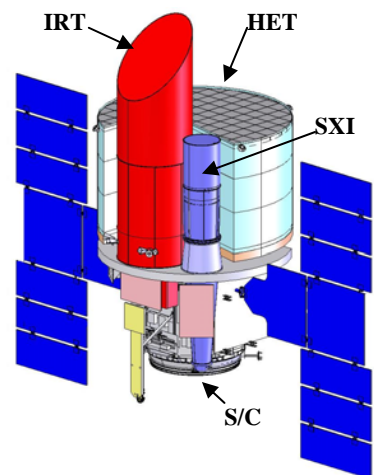

**Fig. 1:** *EXIST* as proposed to Astro2010.

---
[1] jaesub@head.cfa.harvard.edu; phone 1 617 496-7512; fax 1 617 496-7577

supermassive black holes.

The SXI (0.1 – 10 keV) will allow finer source positioning (<2″) for GRBs and transients with X-ray afterglows [2], assisting source identification in the optical/IR images acquired by the IRT. For AGN, the combined broad band coverage (0.1 – 600 keV) between the HET and SXI will unambiguously determine the source type (e.g. AGN type I vs II). The IRT will provide simultaneous imaging and spectroscopic measurements of the target over about 4′ × 4′ field of view (FoV) in four separate bands covering 0.3 – 2.2 μm [3]. The near IR coverage will enable the measurement of highly redshifted Lyα-break out to $z \sim 20$. For bright bursts, the high resolution spectra ($R \sim 3000$) will reveal the local environment of these cosmological events, capturing the state of the early Universe at Epoch of Reionization. During the three year followup pointed mission phase, the IRT will take the image and spectra of thousands of sources, generating a rich data set for various studies.

Here we briefly review the current design concept of the HET (§2), and we report the progress and future plan (§3–5) of the technology development for the HET, which is underway through a series of balloon-borne wide field hard X-ray telescope experiments *ProtoEXIST*. We have completed the first part of the program *ProtoEXIST1* with a successful high-altitude balloon flight in Oct, 2009 (§3.1). Currently we are in the middle of the design and layout work for the readout electronics boards for *ProtoEXIST2 & 3* (§3.2-3.5). See also Grindlay et al (2010) [1] for the *EXIST* mission overview and Hong et al (2009) [4] for a more in-depth overview of the design concept and expected performance of the HET. For the detailed flight performance of the *ProtoEXIST1* telescope, see Hong et al (2010) [5] and for the telescope integration and assembly, see Allen et al (2010) [6,7].

## 2. HIGH ENERGY TELESCOPE (HET)

As demonstrated by the *Swift* and earlier missions, the hard X-ray band is ideal for capturing transient sources because of the low density of the steady background sources and the relatively high X-ray flux of transients allowing precise localization. The HET takes advantage of the design and operational heritage of the *Swift*/BAT, *INTEGRAL*/IBIS and *Fermi*/LAT [8]. The HET is a single wide-field (90º × 70º) hard X-ray (5 – 600 keV) coded-aperture telescope (Fig. 2), which is similar to the BAT but with a much greater sensitivity, a higher angular resolution and a broader energy band coverage. This simple concept of one large coded-aperture telescope is in fact the outcome of an extensive design study effort that explored various options (e.g. multiple small coded-aperture telescopes) to find an optimal configuration under the given mass, volume and power constraints.

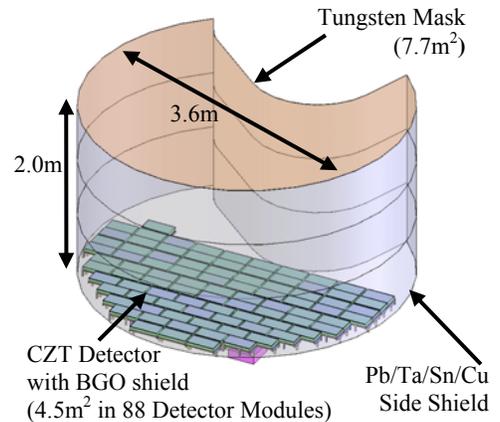

**Fig. 2** The HET design overview and the CZT detector plane.

For X-ray detection, we employ a large array (4.5 m$^2$) of CdZnTe (CZT) detector, which is surrounded by graded-Z passive side shields (Pb/Ta/Sn/Cu) and Bismuth Germanate (BGO) rear anti-coincidence shields (Fig. 2). The large detector plane consists of 88 identical Detector Module (DMs), and each DM is an assembly of 128 identical Detector Crystal Units (DCUs) on a subsequent event processing electronics board with a rear BGO shield. Each DCU is made of a 2×2 cm$^2$ CZT crystal (32 × 32 pixels, 0.6 mm pixel) bonded to an Application Specific Integrated Circuit (ASIC) with a matching 2-D array of 32 × 32 channels. We employ 0.5 cm thick CZT crystals for the broad band coverage (5 – 600 keV) and for GRBs, BGO scintillators (0.7 cm thick) will serve as a high energy calorimeter, extending the energy coverage up to a few MeVs. The CZT detector plane of the HET is hierarchically modular both in mechanical packaging and in the data concentration and processing. Based on the architecture implemented on *Swift*/BAT, the CZT detector allows both redundancy and fast imaging. The Tungsten mask contains pixels of two different scales: fine thin (0.3 mm thick, 1.25 mm pixel pitch) and coarse thick (3 mm thick, 15 mm pixel pitch) elements [9,10]. The two-scale pixel mask allows fine angular resolution (2.4′) and rapid source localization (<20″ in 10 sec) through an efficient two-step imaging algorithm. It also allows the broad band (5 – 600 keV) and wide field coverage (90° × 70°) without auto-collimation.

During the first two years, the HET will operate in the scanning survey mode, continuously sweeping nearly the entire sky every two orbits. The scanning mode will be only interrupted by the followup observations of GRBs and transients (about 3 day$^{-1}$ on average). The continuous scan minimizes imaging noise in coded-aperture imaging and provides a high probability of capturing rare events, while allowing nearly continuous monitoring of variable sources. The survey sensitivity of the HET will reach ~0.06 mCrab in the 5 – 150 keV range in the first two years. In the next three years, *EXIST* will operate in the pointing mode, performing followup observations of the selected sources from ~20,000 AGN detected in the survey mode. The GRB and transient followups will continue throughout the mission.

**Table 1** CZT Detector Plane and Telescope Configurations: *ProtoEXIST* to *EXIST* vs. *Swift*/BAT

| Telescope | # CZTs ($2\times2\times0.5$ cm$^3$) | Det. Area (cm$^2$) | Det. Pixel (mm) | Mask Pixel/Thickness (mm/mm) | Ang. Res. | FoV (50%) (deg) | FE ASIC (Power) |
|---|---|---|---|---|---|---|---|
| *ProtoEXIST1* | 64 | 256 | 2.5 | 4.7 / 3 | 20.3′ | 20º | RadNET (100–150 µW/pix) |
| *ProtoEXIST2* | 16 | 64 | 0.6 | 1.25 / 0.3; 4.7 / 3[a] | 2.4′ | 9.2º | DB-ASIC (50–80 µW/pix) |
| *ProtoEXIST3* | 48 | 192 | 0.6 | | | | EX-ASIC (20 µW/pix) |
| *EXIST*/HET | 11264 | 45000 | 0.6 | 1.25 / 0.3; 15 / 3[a] | 2.4′ | 90º×70º [b] | EXF-ASIC (20 µW/pix) |
| *Swift*/BAT | 31678[c] | ~5000 | 4 | 5 / 2 | 22′ | 100º×70º | XA1.23 (2 mW/pix) |

[a] Hybrid mask  [b] 10% coding  [c] CZT 4×4×1 mm$^3$

## 3. CZT DETECTOR DEVELOPMENT TRHOUGH *PROTOEXIST*

We have been developing the detector technology for the HET through a series of balloon-borne wide-field X-ray telescope experiments *ProtoEXIST1, 2 & 3*. Table 1 compares the telescope parameters of *ProtoEXIST* and *EXIST*/HET in comparison with *Swift*/BAT. The CZT detector in the HET will be ~9 times larger in area with ~44 times higher in pixel density than is in the BAT. We have completed the first part of the program *ProtoEXIST1* with a successful flight in Oct 2009. As of this writing, we are in the middle of the 2nd phase, *ProtoEXIST2*, where we will demonstrate our packing architectures for very fine pixel (0.6 mm) large area detector. In *ProtoEXIST3*, we will achieve the power constraint needed for *EXIST* (20 µW/pix). We also plan to demonstrate their performance in a near space environment through another balloon flight with a focal plane of the combined *ProtoEXIST2* and *3* detector arrays (64 and 192 cm$^2$ each) (§5).

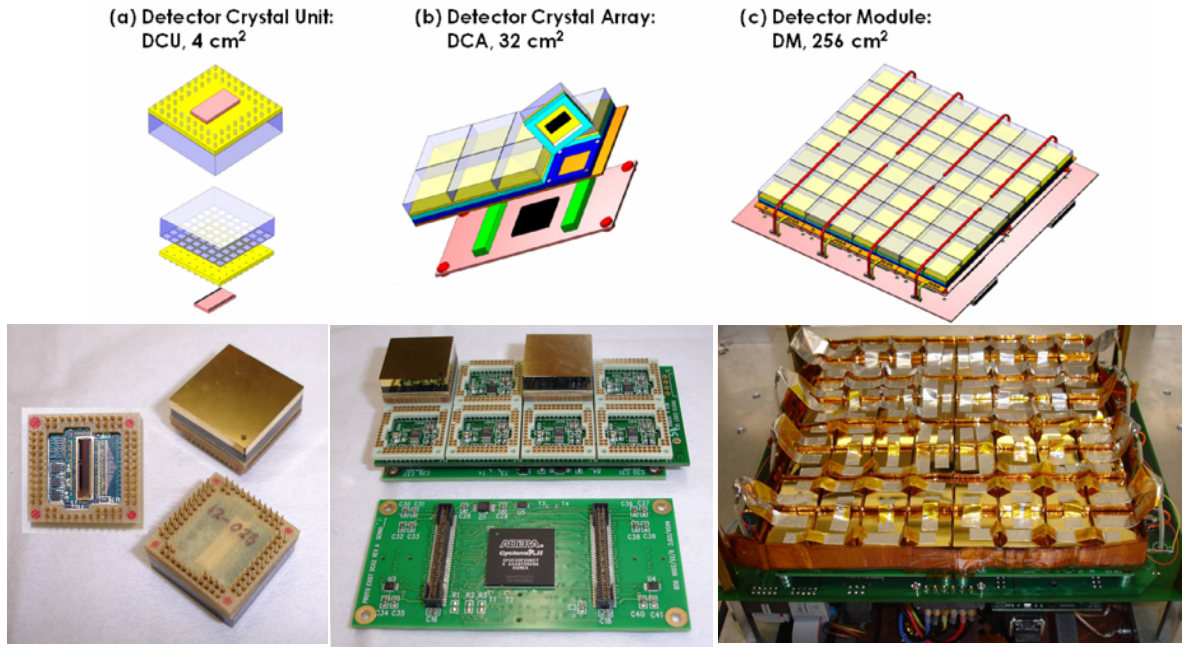

**Fig. 3:** *ProtoEXIST1* detector plane assembly **(a)** Detector Crystal Unit (DCU) consists of $2 \times 2$ cm$^2$ × 0.5 cm CZT bonded via an interposer board to the 1D RadNET ASIC (pink) below. **(b)** Detector Crystal Array (DCA) consists of a 2x4 array of DCUs. **(c)** Final view of flight Detector Module (DM) covers $16 \times 16$ cm$^2$ with a $2 \times 4$ array of DCAs. HV bias (−600V) lines connect to the 4 DCAs on each side.

### 3.1 *ProtoEXIST1* with RadNET ASIC

*ProtoEXIST1* is the first implementation of a close-tiled fine pixel imaging CZT array in our program to develop the technology required for *EXIST*/HET. The initial report on the detector assembly and performance is given in Hong et al (2009) [11]. The full detector plane covers $16 \times 16$ cm$^2$ with $8 \times 8$ CZT crystals, allowing $64 \times 64$ imaging pixels with 2.5 mm pixel pitch. All the 64 CZT crystals are tiled with a uniform gap (~900 μm for $19.5 \times 19.5$ mm$^2$ CZT). This is a factor of four or larger in area than any previous fine-pixel contiguous imaging hard X-ray detectors. The full detector array was integrated into its flight configuration in Sept, 2009 (Fig. 3) [5,6,7].

The basic building block is the Detector Crystal Unit (DCU) that consists of a $2 \times 2$ cm$^2 \times 0.5$ cm CZT crystal with $8 \times 8$ pixels bonded via an interposer board to a RadNET ASIC underneath for imaging readout (Fig. 3a). Next, a $2 \times 4$ array of DCUs are tiled on two vertically stacked electronics boards for a Detector Crystal Array (DCA) (Fig. 3b). The DCA electronics boards contain 4 ADCs and a Field Programmable Gate Array (FPGA) with 4 independent data processing channels (1 ADC and data channel per 2 DCUs). Finally, we closely tile $2 \times 4$ DCAs on a motherboard, the FPGA-Controller Board (FCB), to complete the detector plane or Detector Module (DM; Fig. 3c) for *ProtoEXIST1*. The FCB contains a larger FPGA to process 32 independent data streams (1 per 2 DCUs). A *NetBurner* card is mounted on the FCB to collect the science data from the FPGA and other HouseKeeping (HK) information, and pass them to a flight computer through an Ethernet port.

We obtain CZTs from Redlen Technologies[2], which produces high quality CZTs at low cost through their patented traveling heater CZT growth technique. We bias CZTs at −600V on the cathode using a thin (2 mil) Al tape, which allows easy replacement of DCUs if needed. The Redlen CZTs show < 0.5 nA/pix leakage currents under −600V. Each DCU is surrounded by Kapton-coated Copper shields to minimize the apparent interference between units, and the shields reduce the electronics noise down to the level of a single DCU operation. CZTs were bonded on an interposer board (IPB) with two low-cost, low-temperature ($\leq 100$ C) techniques – low temperature solder bonds from Delphon/QuikPak[3] and Transient Liquid Phase Sintering (TLPS) bonds from Creative Electron Inc[4].

The 2D $8 \times 8$ pixel signals from the CZT are converted through the IPB to the 64 channel 1D input for the RadNET ASIC for *ProtoEXIST1*. The RadNET ASIC, originally developed for a Homeland security project, was chosen for *ProtoEXIST1* mainly because of its relatively low power consumption (~ 100 – 150 μW/pix) and the expected advance of the subsequent ASIC members in the same development family. The low power ASIC is essential for *EXIST*/HET where a large number of pixels (~12M) are needed for fine source positioning and high sensitivity. In addition, the RadNet ASIC provides a wide dynamic range with low electronics noise suitable for the hard X-ray band and allows flexible readout modes – multi-pixel pulse profile readout – required for depth sensing and event reconstruction of multi-pixel triggers arising from charge splitting or Compton scattering. The CZT detectors in a full DCA of *ProtoEXIST1* operating on the ground exhibit ~ 3.2 keV FWHM resolution at 60 keV after a full calibration [4]. The complete detector module shows ~ 4 keV FWHM resolution during the flight after a preliminary calibration (Fig. 4).

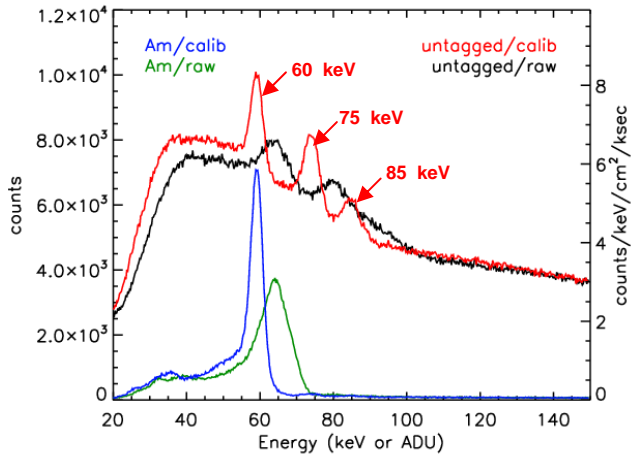
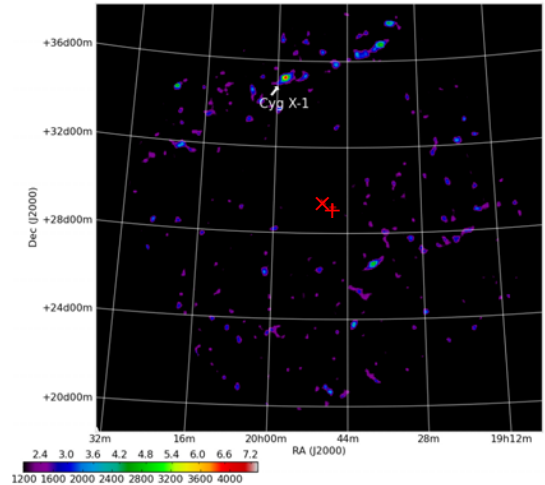

**Fig. 4:** In flight background (calibrated vs. raw) untagged by CsI shield, and $^{241}$Am Calibration source (60 keV) spectra (calibrated, vs. raw). $K_\alpha$ and $K_\beta$ lines (75, 85 keV) from passive Pb side shields (not fully graded) are present (from CR-induced fluorescence) at float. The detector low energy threshold was ~35 keV.

**Fig. 5:** First-light *ProtoEXIST1* image (30−100 keV) of Cyg X-1 and "ghost" image at lower right, due to the 5.6° off-axis pointing and thus partially coded position and cyclic URA. Pointing position (from star camera) is plotted as × and center of X-ray image as + for a boresight offset of 0.81°.

---

[2] http://www.redlen.com
[3] http://www.delphon.com

The *ProtoEXIST1* detector plane was surrounded by partially graded-Z side shields (Pb/Sn/Cu) and a 26 × 26 cm$^2$ × 2 cm CsI rear active shield. The mask was made of 12 identical layers of 0.3 mm thick Tungsten sheets with a pattern of 2 × 2 uniformly redundant array (URA) in 64 × 64 pixels with 4.7 mm pixel pitch. The mask was located at 90 cm above the detector plane, providing 10° × 10° FoV at 50% coding fraction with a 20′ resolution. A 200 nCi $^{241}$Am calibration (AmCal) source was mounted on the side of the passive shield to monitor the gain variation of the CZT detectors during the flight. The detector plane, the CsI rear shield and the lower part of the side shield sit inside of a pressure vessel (PV), where the inside pressure and temperature are maintained at ~1 atm and 20 °C respectively. The mask, its support frame, and the upper part of the side shields sit on the top of the PV. The payload of the *ProtoEXIST1* telescope was assembled into the gondola with a new pointing and aspect system including a daytime star camera in late Sept, 2009.

After six failed attempts due to the bad weather, we had a successful launch of the *ProtoEXIST1* telescope on Oct. 9, 2009. The flight had duration 6.5 hr at 131,000 ft. The overall performance of the X-ray telescope including the CZT detectors was excellent. We accomplished all three main objectives, which were 1) to measure detector background rates and overall performance in near-space conditions; 2) to verify thermal, power, and overall detector functions, and 3) to obtain a "first light" image on a bright (> 100 mCrab) source. Background rates were stable over the flight with ~4 × 10$^{-3}$ cts cm$^{-2}$ sec$^{-1}$ keV$^{-1}$ at ~100 keV and roughly as predicted. High-Z cosmic rays and subsequent charge-saturation did not upset the detector or FPGAs or computer systems. Both the raw and calibrated spectra of the background along with the onboard AmCal events are shown in Fig. 4. The insulation and thermal system of the pressure vessel maintained the detector plane at ~20 ºC throughout the flight.

During the short flight, the only available celestial source brighter than ~100 mCrab (detectable by our 256 cm$^2$ wide-field detector in ~1 hour) was Cyg X-1 at the very end of the flight. About an hour observation of the source with stable pointing only for the final 10 min produced the 7.2σ detection at ~6.5º off the center of the FoV (Fig. 5). The offset was due to a failure of the Differential Global Positional System (DGPS) system which required use of the uncalibrated onboard magnetometer for target acquisition.

### 3.2 *ProtoEXIST2* with DB-ASIC

Two main aspects needed for the HET CZT detectors in comparison with the *ProtoEXIST1* CZT detectors are the smaller pixel size and lower power consumption (Table 1). In *ProtoEXIST2*, we enable 0.6 mm pixel imaging using the Direct Bond (DB) ASIC developed for *NuSTAR* (Fig. 6) [12]. The form factor of the DB-ASIC allows a direct bond between an ASIC and a CZT crystal with the matching pattern of the CZT anode pixels and ASIC input pads (a 2D 32 × 32 array), which is essential for such a high pixel density detector. Belonging to the same ASIC family line as the RadNET ASIC, the DB-ASIC also allows for the same critical features needed for *EXIST*/HET such as multi-pixel pulse profile readout with 16 sampling capacitors per pixel. In addition, there are quite a few improvements in the signal processing circuitry in the DB-ASIC. Consequently the electronics noise of the DB-ASIC is substantially lower (~0.34 keV FWHM) than the same (~2.0 keV FWHM) of the RadNET ASIC (Fig. 7). The DB-ASIC also includes an on-chip analog-digital converter (ADC), which minimizes the risk of the signal degradation during the analog-digital conversion and simplifies the subsequent readout electronics.

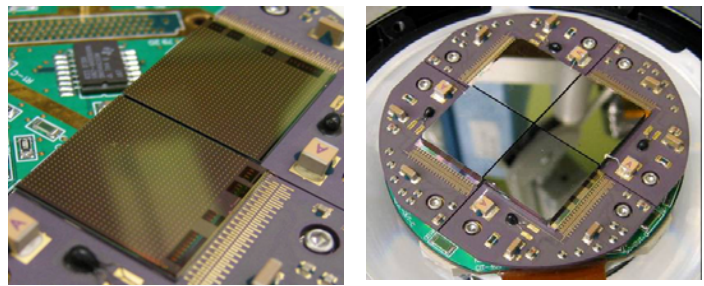

**Fig. 6:** ASIC and CZT tiling for *NuSTAR* as model for *ProtoEXIST2*. (Left) two DB ASICs mounted on carrier board for *NuSTAR*, each with 32 × 32 array of pixel contacts on upper surface and (Right) 2 × 2 close-tiled array of CZT crystals (20.2 × 20.2 mm$^2$) each bonded to its ASIC. The control, data and power inputs are accessible via pads on the upper surface of the ASIC as well (e.g. on the right side of the nearest ASIC shown) and are directly wire bonded to the carrier board. The geometry of the complete *NuSTAR* detector plane motivates the DCA for *ProtoEXIST2* (Fig. 3.3b).

---

[4] http://www.creativeelectron.com

In *ProtoEXIST2*, we will develop the large area of CZT tiling architecture with DB-ASIC based on our experience with *ProtoEXIST1*. The readout system and event processing boards will allow both *ProtoEXIST2* and *3* modules, so that a combined *ProtoEXIST2-3* detector and telescope system, with a shared coded mask, will ultimately be tested in a proposed balloon flight to do high resolution hard X-ray imaging and comparative detector and telescope tests in a near-space environment (see §5).

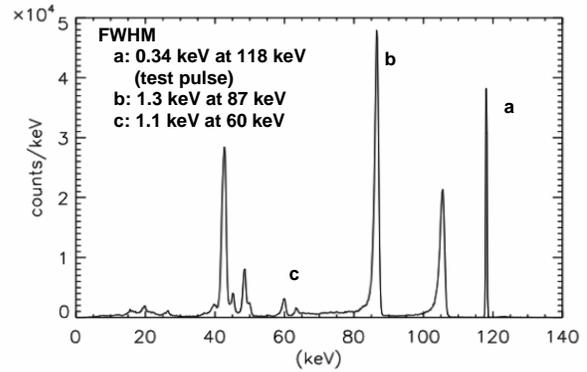

**Fig. 7:** An energy spectrum taken with an eV Products crystal read out with a DB-ASIC. The energy resolution is 1.3 keV (FWHM) at 86.5 keV and electronics noise of test pulse is 0.34 keV (FWHM) at 118 keV.

The *ProtoEXIST2* detector modularization (Fig. 8) begins with a single DCU, where a CZT is directly flip-chip bonded onto a DB-ASIC, which sits on an ASIC Carrier Board (ACB, Fig. 8a). The wirebond input/outputs to the DB-ASIC require a ~5 mm "extension" on one side of the ASIC, which means that near gapless tiling (single pixel gap) can be achieved only with a 2 × 2 array of CZT crystals, as shown in Fig. 8b. This close tiled 2 × 2 array of DCUs forms a Detector Crystal Array (DCA, Fig. 8b), and 2 × 2 such DCAs constitute a Quad Detector Module (QDM, Fig. 8c) that is readout by an Event Logic Board (ELB). Four such QDMs are in turn mounted on the Detector Module Board (DMB) to constitute the full Detector Module (DM, Fig. 8d). *ProtoEXIST2* is thus a 3 board system (ACB, ELB and DMB) and is more compact than *ProtoEXIST1* (4 boards), which allows better shielding for less structural mass.

Each ACB is designed to accommodate a single 5 mm thick 2 × 2 cm$^2$ CZT crystal directly bonded to a single *NuSTAR* DB-ASIC. The design of the *ProtoEXIST2* ACB is based heavily on the design of the *NuSTAR* carrier board and test board (Fig. 6 & 9). The ACB is primarily used to provide power to and route signals from the DB-ASIC. The form factor of the DB-ASIC matches with a CZT with 32 × 32 pixilated anode with a pixel pitch of 0.6 mm and an optional 0.16 mm wide "guard ring" which is held at a constant –4 V potential and extends around the perimeter of the CZT anode. A 4 × 4 array of ACBs are mounted on the ELB which contains the electronics necessary for the preliminary processing of event data collected from the 16 ACBs. Preliminary processing is carried out in 16 complex programmable logic devices (CPLD) which are mounted to the ELB and each matched with an individual ACB. Once preliminary processing is complete the refined event data is transferred from the 4 ELBs to the next layer down, the detector module board (DMB). Here GPS timing information, event coincidence tags from the calibration source and active shielding system are applied to the data which is then collated and formed into packets for transmission via an ethernet interface to the flight computer for storage and analysis. Each ELB is handled by an individual FPGA located on the DMB and the collation and output to the ethernet interface card (Netburner) is handled by a single FPGA.

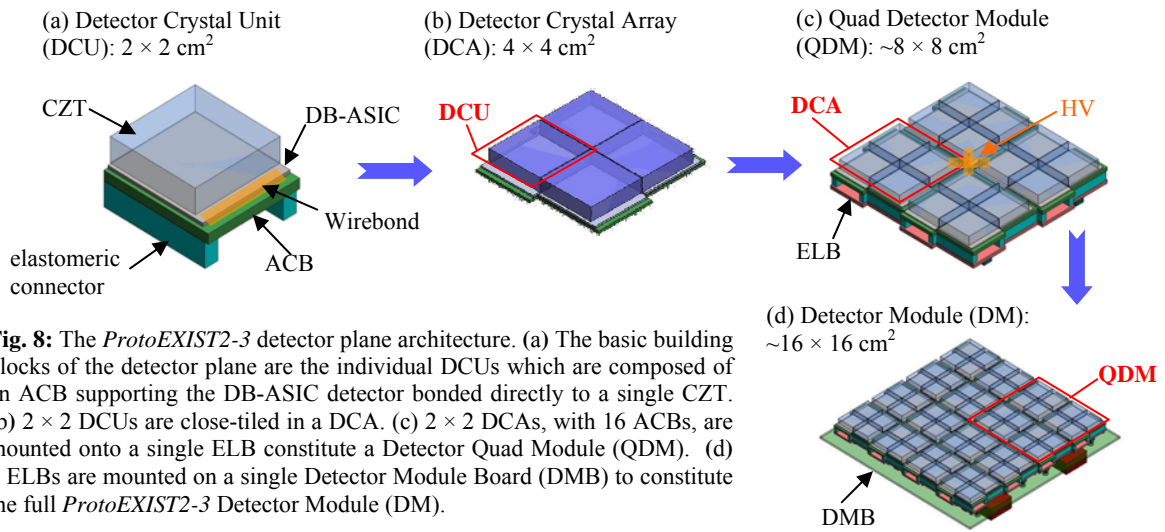

**Fig. 8:** The *ProtoEXIST2-3* detector plane architecture. (a) The basic building blocks of the detector plane are the individual DCUs which are composed of an ACB supporting the DB-ASIC detector bonded directly to a single CZT. (b) 2 × 2 DCUs are close-tiled in a DCA. (c) 2 × 2 DCAs, with 16 ACBs, are mounted onto a single ELB constitute a Detector Quad Module (QDM). (d) 4 ELBs are mounted on a single Detector Module Board (DMB) to constitute the full *ProtoEXIST2-3* Detector Module (DM).

The ELB and the DMB are designed to preserve pixel pitch across the entire detector plane while minimizing the separation distance between individual DCUs. In order to achieve this, a two scale tiling approach was adopted. At the smaller scale the DCAs have been adapted to mimic the tiling pattern utilized by *NuSTAR* leaving one pixel gap between the edge pixels of neighboring CZT crystals in the DCA. On the larger scale a "jigsaw" layout of DCAs which interleaves the excess board area and has been adopted resulting in a ~5 mm gap between the nearest CZT crystals of neighboring DCAs. The 4 ELBs are mounted on the DMB such that the uniformity of the large scale tiling as well as pixel pitch is preserved across the entire $17.6 \times 17.6$ cm$^2$ detector plane. The CZT pixels are maintained within ~50 μm throughout the detector plane by the precise alignment of CZT and ASIC using an optical camera (<25 μm precision) during the bonding process, the similar procedure for the ASIC to ACB bonding, and alignment pins between the three boards.

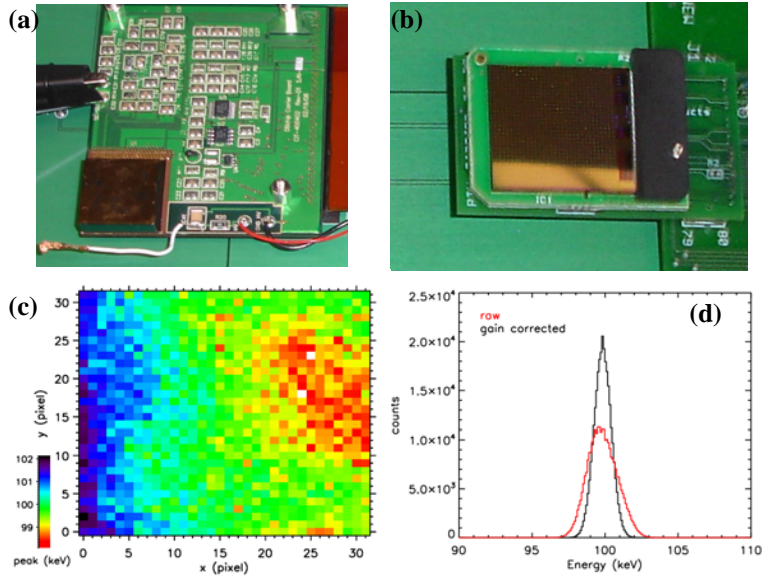

**Fig. 9:** (a) An early version of a DCU with the DB-ASIC (b) A bare DB-ASIC test setup. (c) The gain variation across a DB-ASIC of the pulser. (d) Energy spectra taken from a pulsar data set, which exhibits ~1 keV FWHM noise. The relatively large noise is dominated by the external input voltage for the pulse height control, and the internal ASIC noise is ≤0.3 keV.

A −600 V bias is applied between the CZT cathode and anode pixels by bringing 4 HV leads up through the center of each ELB (Fig. 8c) that are each connected to a single DCA. The cathodes of the individual DCUs in a single DCA are coupled via conductive Al tape, as in *ProtoEXIST1* (Fig. 2c). A RC filter in close proximity to the termination point of each HV lead ensures minimal noise. Two EMCO low noise high voltage power supplies are mounted on the bottom of the DMB and each provide the bias for 8 HV leads. Command and control of the *ProtoEXIST2* DM as well as data acquisition uses an internal private network via an ethernet interface implemented using a MOD 2582 *Netburner* on the underside of the DMB, and driver code developed for the *ProtoEXIST1* command, control and DAQ.

### 3.3 *ProtoEXIST3* with EX-ASIC

*ProtoEXIST3* will be the final implementation of the advanced fine pixel imaging CZT array, meeting all the technology requirements for *EXIST*/HET (Table 1). The key remaining *required* aspect of the CZT detectors after *ProtoEXIST2* is the ASIC with lower power consumption (≤ 20 μW/pix). The combination of the large area (4.5 m$^2$) and the fine pixels (0.6 mm) of the CZT detectors required for precise (<20″, 90% confidence locations) positions of GRBs and transients by *EXIST*/HET demands a readout system with a large number of electronic channels or pixels (~12M). The power constraint for the Front End Electronics (FEE) requires low-power EX-ASICs (~20 μW/pixel). It is not easy to increase the FEE's power budget for *EXIST*/HET since enlarging solar panels (and the S/C battery) would greatly drive up mission complexity – e.g. moments of inertia would increase, requiring even more massive and high power reaction wheels. The larger solar panel sizes also would result in larger fairing size for the launcher, further driving up mission cost.

The main development route to a lower power ASIC is a modification of the current DB-ASIC to the *EXIST*-Specific EX-ASIC. The lower power consumption by the required factor of ~4 will mean a decrease in energy resolution by a factor of ~2 ($P \sim I^2R$, and detector noise scales with $I$). Based on the on-going improvements in this ASIC family (e.g. ~50 – 80 μW/pix for DB-ASIC vs. ~100 – 150 μW/pix for RadNET ASIC, Table 1) and a relatively forgiving noise requirement (~1.2 keV FWHM) for *EXIST*/HET in comparison with *NuSTAR* (<0.4 keV), we expect the modification for further power reduction is rather straightforward, but clearly we need to demonstrate this and to assess the low energy threshold (5 keV) and energy resolution changes. Since both DB-ASIC and EX-ASIC will share the nearly identical form factors, we can employ the same three board systems for integrating the EX-ASIC based DCUs except for a small modification of ACB needed for flip-chip bond in place of wirebond. The combined *ProtoEXIST2-3* detector plane will consists of 1 DB-ASIC based QDM and 3 EX-ASIC based QDMs with an option of 1 EXF-ASIC based QDM (§3.4) as shown in Fig. 10.

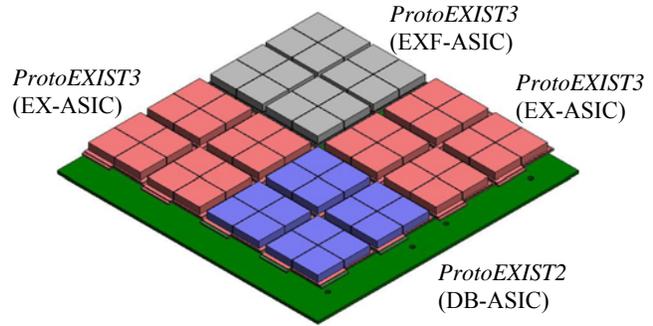

**Fig. 10:** The combined ProtoEXIST2-3 detector plane will consist of 1 DB-ASIC based QDM and 2 or 3 EX-ASIC based QDMs with an option of 1 EXF-ASIC based QDM.

### 3.4 *ProtoEXIST3* with EXF-ASIC

An additional *desirable* aspect of the new ASIC is the form factor that eliminates the need for wirebonds, allowing the most efficient packaging and assembly. We plan to pursue this through the new emerging technology of *microvias*. This will result in the development of a "final", or "flight" version of the EX-ASIC, namely the EXF-ASIC. Although not required, minimizing gaps between detectors is important since a) it allows the full detector area available to be used (the fractional area lost, with gaps, is ~20% and b) it allows reduced background in otherwise edge-illuminated detector pixels. The (near-) gapless DCA tiling would enable a 30% reduction in background at 30–100 keV for edge pixels as estimated by the observed higher background counts in the edge pixels during the *ProtoEXIST1* flight [6]. This, in turn, allows the detector plane images to be more uniform, thereby decreasing systematic noise in the imaging process.

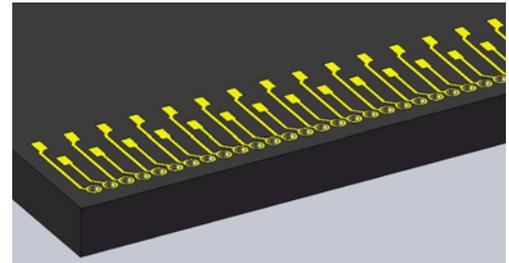

**Fig. 11:** Schematic model of the bottom side of the EXF-ASIC utilizing microvia technology. Shown above are the 6 mil diameter microvias attached to input pads which are flip chip bonded to the EXF carrier board on which the EXF-ASIC is mounted.

Near-gapless tiling can be achieved by eliminating the wire bond pads on the DB chip and the corresponding need for an extension of the chip to bond to the ACB (see Fig. 8a). The total "protrusion" of the wirebonding "jigsaw" is ~5 mm. This can be reduced to only ~1mm if instead of wirebonds, all the 87 input/output connections to the EX-ASIC are replaced by microvias, which connect the top (CZT side) of the ASIC directly to the ACB below by flip chip bonding, as shown in Fig. 11. In Fig. 12 we show a single QDM tiled with the EX-ASIC and its required gaps between DCAs vs. with the proposed EXF-ASIC tiling design.

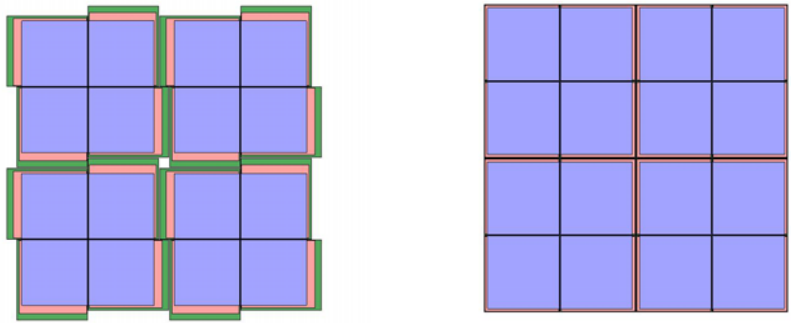

**Fig. 12:** *ProtoEXIST3* ¼ detector module (or QDM) if tiled, *left*, with the EX-ASIC and gaps due wirebond pads on the ASIC (red) and ACB (green) vs. if tiled, *right*, with the EXF-ASIC enabled by microvias replacing the bond pads on the ASIC to couple to the ACBs below. CZTs in blue.

The microvias are laser drilled in the outer 1 mm of the Si wafer that will form the 20.06 mm x 21 mm EXF chip (vs. the DB or EX ASIC which has dimensions $20.06 \times 21.79$ mm$^2$). The larger gap (~5 mm) is due to the additional size for the ACB bonding pad area, with two rows of 43 bond pads, as well as the microvias that must straddle both rows of bonding pads to bring the signals down to the ACB. The laser drilling would be the very first step in the ASIC production process so that it does not interfere with the production of the ASIC itself. In order for such fine (6 mil) holes to be reliably drilled, and then given conductive inner walls to become a conducting via, the Si wafer thickness must be comparable in thickness to the hole size and so be ~250 μm vs. the 750 μm thickness on the present DB (and thus also, the EX) chip.

Our early research on the plausibility of the microvia technology for *ProtoEXIST3* and HET is very promising. High demand of compact circuitry for many commercial applications is expected to provide additional boost on the maturity and low-cost availability of the technology.

## 4. OTHER DEVELOPEMENT

### 4.1 Mask Design

As in *EXIST*/HET, *ProtoEXIST2-3* will utilize a hybrid mask in order to test the proposed imaging scheme for the *EXIST* mission which requires the rapid and efficient localization of hard X-ray sources from a scanning platform, where images must be generated continuously on short (< 2 sec) timescales and co-added. A hybrid coded aperture mask is the superposition of a fine pitch mask and a coarse pitch mask, where each open pixel of the coarse pattern reveals the fine pitch of the fine mask [11].

The mask is constructed of multiple layers of thin laminated tungsten sheets, each 0.3 mm thick, with an identical random mask pattern etched into the surface. For the construction of a hybrid mask, a single tungsten sheet with a small pitch (1.25 mm) random pattern is etched into the surface. This sheet is then placed in the center of the mask stack such that the smaller scale pattern overlays the large scale pattern. Both the large and small scale patterns have a 0.5 open fraction, which leaves a 0.25 total open fraction for high resolution imaging. In *ProtoEXIST2-3*, the large scale pixels will utilize the original *ProtoEXIST1* URA mask pattern with pixel pitch of 4.7 mm, with an additional small scale mask introduced with a pixel pitch of 1.25 mm, which will enable us to test inflight the imaging subroutines required for the *EXIST* mission.

### 4.2 Active Shield using Avalanche photodiode

We plan to employ BGO scintillators underneath the CZT detectors in *EXIST*/HET in order to reduce the atmospheric albedo and cosmic-ray induced background. In order to explore the performance of the BGO shield with CZT detectors, in *ProtoEXIST2-3*, we will use a BGO scintillator ($18 \times 18$ cm$^2$ $\times$ 0.7 cm) read out by two light guides (on opposite edges), each coupled to two Avalanch Photo Diodes (APDs). The *ProtoEXIST2-3* BGO shield will be more tightly packaged and closer to the *ProtoEXIST1* detector plane than was possible with the CsI shield for *ProtoEXIST1*. The proximity of the CZT array and BGO scintillator is important to minimize the internal background induced by cosmic-ray interaction with the material between the CZT and BGO crystals. The use of APDs eliminates the bulky PMTs and enables a more compact ($8 \times 8$ mm$^2$ APD, well matched to the 7 mm thick BGO) and higher quantum efficiency shield. Since the *ProtoEXIST2-3* detector is more compact than *ProtoEXIST1* (3 vs. 4 boards and flat connectors), it would require a more efficient heat transfer to the PV cold frame from the FPGAs on the DMB.

The *ProtoEXIST2-3* flight will produce the data set that can be used to compare two shield systems: *ProtoEXIST1*/CsI and *ProtoEXIST2-3*/BGO (Table 2). The results will be compared to the GEANT4 based background simulation, and will be used to optimize the shield design (e.g. thickness) for *EXIST*/HET.

**Table 2** Proposed Balloon flight configurations of the *ProtoEXIST* telescopes

| Parameters | *ProtoEXIST2-3* Flight |
|---|---|
| Flight Schedule | September, 2012 |
| Focal length | 200 cm |
| FoV | 9.2º × 9.2º (50% coding) |
| Payload | *ProtoEXIST1*: 256 cm$^2$ CZT with CsI shield <br> Det. Pix: 2.5 mm, Mask Pix: 4.7 mm, Ang. Res: 9.2′ <br> *ProtoEXIST2-3*: 64 cm$^2$ *ProtoEXIST2* CZT <br> + 192 cm$^2$ *ProtoEXIST3* CZT with BGO shield <br> Det. Pix: 0.6 mm <br> Fine Mask Pix: 1.25 mm, Ang. Res: 2.4′ (E ≤ 60-100keV) <br> Coarse Mask Pix: 4.7 mm, Ang. Res: 8.1′ (E ≥ 100 keV) |
| Primary goals | • Test the 2–3$^{nd}$ generation adv. CZT imaging detector <br> • High resolution imaging, with scanning <br> • Compare CsI vs. BGO active shields |

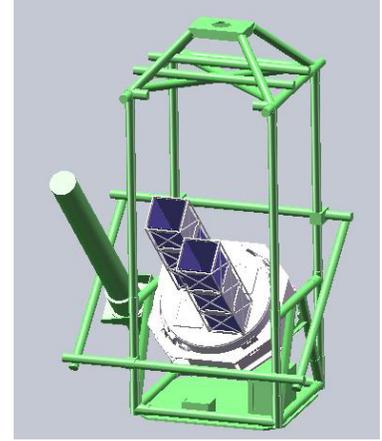

**Fig. 13:** The *ProtoEXIST2-3* flight configuration, as well as *ProtoEXIST1* (upper left) with its focal length extended to 2 m.

## 5. FLIGHT PLAN

Table 2 summarizes the next flight plan and payload configuration (Fig. 13) to test both *ProtoEXIST1* and *2-3* telescopes. Our proposed flight would then compare the imaging and shielding performance of *ProtoEXIST2* (1 QDM) along side *ProtoEXIST3* (3 QDMs) in a single DM. Due to the random mask, and tiled detector planes of each, both can image independently or combined. Each would have the full 2 m focal length proposed for *EXIST*, and thus the same 2.4′ angular resolution, but given mass and gondola constraints would have much narrower FoV. Re-flying *ProtoEXIST1* would allow direct comparison of the diffuse vs. internal background since its FoV is now ~1/4 of what it was on the *ProtoEXIST1* flight. And with the possible addition of a complete DCA with the candidate EXF-ASIC, we would compare the performance of near-gapless tiling with minimum edge pixel.